\documentclass[11pt,a4paper,epsfig]{article}

\usepackage{amsfonts}
\usepackage{epsfig}

\title{\textbf{Composite solitary waves in three-component scalar field theory:
II. Three-body low-energy scattering}}
\author{A. Alonso Izquierdo$^{(a)}$ and J. Mateos Guilarte$^{(b)}$
\\ {\normalsize {\it $^{(a)}$ Departamento de Matematica
Aplicada}, {\it Universidad de Salamanca, SPAIN}}\\{\normalsize
{\it $^{(b)}$ Departamento de Fisica Fundamental and IUFFyM}, {\it
Universidad de Salamanca, SPAIN}}}
\date{}

\begin{document}

\maketitle

\begin{abstract}
We discuss time evolution of some solitary waves described in the
first part of this work. The adiabatic motion of the non-linear
non-dispersive waves composed of three lumps is interpreted as
three-body low energy scattering of these particle-like kinks.
\end{abstract}

\section{Introduction}

After a brief summary of the stability properties of the solitary
waves discovered in the first part we shall devote this second
part to the analysis of low energy dynamics. Study of Manton's
adiabatic motion of solitary waves \cite{Manton} is very simple in
this case because the geometric metric of the moduli (parameter)
space that governs the dynamics does not depend on the
translational parameter $x_0$; thus, it is always possible to
apply a transformation that leads to an Euclidean metric. On the
other hand, there are only a few works that have addressed
research into solitary waves in three-component scalar field
theory models, see \cite{Aai3,Aai4,Aai5,BLW}. In this richer case
there are three-parametric families of kinks, and the adiabatic
evolution of three-body lumps is associated with metrics with
curvature. The low-energy dynamics of, in this case, three-body
solitary waves is therefore much more intricate.

In this paper, we shall address a three-component scalar field
model that generalizes the two-component model discussed in
Reference \cite{modeloa} to three fields. The organization of the
paper is as follows: In Section 1 some remarks about the stability
of these solutions are given. Finally, in Section 2 we shall
describe the adiabatic evolution on the moduli space of solutions
of configurations composed of three basic lumps.

\section{Some remarks about the stability of solitary waves}

In order to analyze the stability of the solitary wave solutions,
we shall apply different procedures. The direct approach is study
of the spectrum of the second-order fluctuation (Hessian) operator
--formula (10) in part I--, now around kink solutions. A
non-negative spectrum of the kink Hessian operator ensures that a
particular solitary wave solution is stable. Zero eigenvalues in
the spectrum, zero modes, show that certain perturbations send
static solutions into static solutions. Thus, zero modes are
associated to neutral equilibrium directions in a sub-space of the
configuration space where the adiabatic evolution of solitary
waves takes place. It is easy to check that the existence of a
kink family $\bar\phi(x,c)$ bears the presence of an eigenfunction
$\frac{\partial \bar\phi}{\partial c}$ that belongs to the kernel
of ${\cal H }$, i.e., $\frac{\partial \bar\phi}{\partial c}$ is a
zero mode. When detailed information about the spectrum of the
Hessian is lacking, an interesting procedure to study the
stability of kink families is the application of Morse index
theorem. A point --referred to as a conjugate or focal point--
that is crossed by all the members of this family implies the
existence of a negative eigenvalues in the spectrum of the Hessian
operator \cite{Ito2,J1}.

\subsection{Stability of basic solitary waves}

Let us consider first basic solitary waves, i.e., the $K_1^{OB}$,
$K_1^{CD}$ and $K_1^{BC}$ kinks.

\begin{enumerate}

\item  $K_1^{OB}$ kink. Because only the third component is non-null,
the Hessian is a diagonal $3\times 3$-matrix differential
operator with entries:
\begin{eqnarray*}
{\cal H}_{11}
&=&-\frac{d^2}{dx^2}+1+\bar{\sigma}_3^4+(1-\sigma_3^4) \tanh
\sqrt{2} \bar\sigma_3^2 \bar{x}-\frac{3}{2} \bar\sigma_3^4 {\rm
sech}^2
\sqrt{2} \bar\sigma_3^2 \bar{x}  \\
{\cal H}_{22} &=&-\frac{d^2}{dx^2}+1-2\sigma_2^2
\bar\sigma_2^2-\sigma_3^2(2\sigma_2^2-\sigma_3^2)+\bar\sigma_3^2
(1+\sigma_3^2-2\sigma_2^2) \tanh \sqrt{2} \bar\sigma_3^2 \bar{x}-
\frac{3}{2} \bar\sigma_3^4 {\rm sech}^2
\sqrt{2} \bar\sigma_3^2 \bar{x} \\
{\cal H}_{33} &=&-\frac{d^2}{dx^2}+\frac{1}{2} \bar\sigma_3^4
(10+6 \tanh \sqrt{2} \bar\sigma_3^2 \bar{x}- 15 \, {\rm sech}^2
\sqrt{2} \bar\sigma_3^2 \bar{x}) \qquad .
\end{eqnarray*}
The spectral problem is fully solvable because we are dealing with
three ordinary Schr\"odinger operators of the Posch-Teller type.
${\cal H}_{11}$ and ${\cal H}_{22}$ govern orthogonal fluctuations
to the kink orbit in internal space, whereas ${\cal H}_{33}$ takes
into account fluctuations tangent to the orbit. There are no
discrete eigenvalues in the spectrum of ${\cal H}_{11}$ and ${\cal
H}_{22}$, all the eigenfunctions belonging to the continuous
spectrum. In the first case, the continuous spectrum starts at the
threshold $\frac{\sigma_3^2}{\bar\sigma_3^4}$; $\omega^2
(q)=q^2+\frac{\sigma_3^2}{\bar\sigma_3^4}$ is a non-degenerate
eigenvalue in the
$[\frac{\sigma_3^2}{\bar\sigma_3^4},\frac{1}{\bar\sigma_3^4}]$
interval but it is doubly degenerate for higher values of
$\omega^2$. In the second case, the threshold is
$\frac{\sigma_{32}^4}{\bar\sigma_3^4}$ and the doubly degenerate
spectrum starts at $\frac{\bar\sigma_2^4}{\bar\sigma_3^4}$. On the
other hand, $\frac{\partial \phi}{\partial x}$ is a discrete
eigenfunction of ${\cal H}_{33}$ with $\omega^2=0$ eigenvalue
(zero mode) due to spontaneous breaking of translation symmetry by
the kink solution. The continuous spectrum $\omega^2=q^2+2
\bar\sigma_3^4$ is non-degenerate for $\omega^2 \in
[2\bar\sigma_3^4,8\bar\sigma_3^4]$ and doubly degenerate for
$\omega^2\geq 8\bar\sigma_3^4$. In conclusion, the lack of
non-negative eigenvalues allows us to claim the stability of this
solution.

The Hessian for any other solitary wave solution is a non-diagonal
matrix operator and the task of solving the spectral problem
becomes hopeless. There is, however, a loophole to avoid this
problem (not of necessary use in the $K_1^{OB}$ case). On the
$\phi_3$ axis, the Hamilton characteristic function --formula (25)
in part I-- becomes:
\[
W(\phi_3)={1\over\sqrt{8}}\left[(\phi_3^2(x)-1)^2+2\sigma_3^2\phi_3^2(x)\right]
\qquad .
\]
The $K_1^{OB}$-kink orbit is a flow line of the gradient of $W$, a
polynomial function. Thus, the energy is given by
$|W(\bar{\phi}^O)-W(\bar{\phi}^B)|$, a topological bound, the
absolute minimum of $E$ in the ${\cal C}^{OB}$ topological sector.
Therefore, $K_1^{OB}$ is stable.

\item $K_1^{CD}$ kink. In this case the Hessian is not diagonal
but the alternative argument works exactly in the same manner.
--formula (25) in part I-- now reduces to:
\[
W(\phi_1,\phi_2)={1\over\sqrt{8}}\left[(\phi_1(x)+\phi_2^2(x)-1)^2+2\sigma_2^2\phi_2^2(x)\right]
\qquad .
\]
Again, $|W(\bar{\phi}^C)-W(\bar{\phi}^D)|$ is the absolute minimum
of $E$ in the ${\cal C}^{CD}$ topological sector and the
$K_1^{OB}$ kink is stable.

\item $K_1^{BC}$ kink. The same argument also works with
\[
W(\phi_2,\phi_3)={1\over\sqrt{8}}\left[(\phi_2(x)+\phi_3^2(x)-1)^2+2\sigma_2^2\phi_2^2(x)+2\sigma_3^2\phi_3^2\right]
\]
and $E(K_1^{BC})=|W(\bar{\phi}^B)-W(\bar{\phi^C})|$, guaranteeing
the stability of $K_1^{BC}$.
\end{enumerate}

\subsection{Stability of two-body solitary waves}

The general Hessian operator for $K_2^{OC}(b)$ and $K_2^{BD}(b)$
kinks is a non-diagonal $3\times 3$ matrix differential operator
with an unknown spectrum. It is possible, however, to apply the
Morse index theorem, see \cite{Ito2,J1}. Note in Figures 5(c) and
6(c) that there are no conjugate points -points where all the
trajectories meet- in the congruence of kink trajectories of this
type; the Morse index theorem relating the number of conjugate
points to the number of negative eigenvalues of the Hessian
reveals that both the $K_2^{OC}(b)$ and $K_2^{BD}(b)$ families are
formed by stable kinks. The absence of conjugate points is due to
the fact that the reduced Hamilton characteristic functions
--formula (25) in part I-- for both kinds of solitary waves are
regular functions with no conical singularities where the gradient
flow of $W$ would be undefined.

This result is confirmed directly for a particular member,
$K_2^{OC}(0)$, of the first family. In this case the Hessian
operator --formula (10) in part I-- is diagonal, with the
following entries:
\begin{eqnarray*}
{\cal H}_{11} &=& -\frac{d^2}{dx^2}+1+\sigma_2^4+(1-\sigma_2^4)
\tanh \sqrt{2} \bar\sigma_2^2 x-\frac{3}{2} \bar\sigma_2^4 \,{\rm
sech}^2
\sqrt{2} \bar\sigma_2^2 x \\
{\cal H}_{22} &=&-\frac{d^2}{dx^2}+\frac{1}{2} \bar\sigma_2^4
(10+6 \tanh \sqrt{2} \bar\sigma_2^2 x- 15  {\rm sech}^2
\sqrt{2} \bar\sigma_2^2 x) \\
{\cal H}_{33} &=&-\frac{d^2}{dx^2}+1-2\sigma_3^2
\bar\sigma_3^2-\sigma_2^2(2\sigma_3^2-\sigma_2^2)+\bar\sigma_2^2
(1+\sigma_2^2-2\sigma_3^2) \tanh \sqrt{2} \bar\sigma_2^2 \bar{x}-
\frac{3}{2} \bar\sigma_2^4 {\rm sech}^2 \sqrt{2} \bar\sigma_2^2
\bar{x}
\end{eqnarray*}
and the analysis of the spectral problem is as follows: there are
no discrete eigenvalues in the spectrum of ${\cal H}_{11}$. The
continuous spectrum $\omega^2
(q)=q^2+\frac{\sigma_2^2}{\bar\sigma_2^4}$ starts at the threshold
$\frac{\sigma_2^2}{\bar\sigma_2^4}$ and is non-degenerate in the
$[\frac{\sigma_2^2}{\bar\sigma_2^4},\frac{1}{\bar\sigma_2^4}]$
interval but doubly degenerate for higher values. $\frac{\partial
\bar\phi}{\partial x}$ is a zero mode of ${\cal H}_{22}$ due to
spontaneous symmetry breaking of translational invariance by the
solitary wave. The continuous spectrum $\omega^2=q^2+2
\bar\sigma_2^4$ is non-degenerate for $\omega^2 \in
[2\bar\sigma_2^4,8\bar\sigma_2^4]$ and doubly degenerate for
$\omega^2\geq 8\bar\sigma_2^4$. Finally, the third component,
${\cal H}_{33}$, has a zero mode with eigenfunction
$\frac{\partial \bar{\phi}}{\partial a}(0)$ due to motion without
energy cost in the space of kink orbits. We find a continuous
spectrum, which is non degenerate in the range
$[\frac{\sigma_{32}^4}{\bar\sigma_2^4},\frac{\bar\sigma_3^4}{\bar\sigma_2^4}]$
and doubly degenerate from this value. Again all the eigenvalues
are non-negative and the stability of this solution is
established.

\subsection{Stability of three-body solitary waves}

The $K_3^{OD}(a,b)$ kinks, encompassing the three basic kinks with
no static forces between them, form a set of distinguished
solitary waves of this system. One- and two-body solitary waves
arise in this model because of the embedding of lower dimensional
systems of the same type. They are generic in the sense that this
variety of solutions depends on as many parameters -two apart from
the center of the kinks- as possible for a three-component scalar
field theory. Moreover, the stability of this two-parametric kink
family is easily established by a quick and visual application of
the Morse index theorem. Away from the vacuum points,
$\bar{\phi}^O$ and $\bar{\phi}^D$, the $K_3^{OD}(a,b)$ kink orbits
do not intersect; there are no conjugate points, see Figure 7(c).
Being the gradient flow lines of $W$ --formula (25) in part I--,
conjugate points cannot exist because $W$ is a regular function
and the flow is well defined. Therefore, these solutions are
stable.

Although the Hessian operator is a very complicated $3\times 3$
non-diagonal matrix differential operator and although the
spectrum is essentially unknown, it is easy to prove the existence
of three zero modes: 1) $\frac{\partial\bar{\phi}_a}{\partial
x}(x;a,b)$, 2) $\frac{\partial\bar{\phi}_a}{\partial a}(x;a,b)$,
and 3) $\frac{\partial\bar{\phi}_a}{\partial b}(x;a,b)$. Given a
one-parametric family of solutions $\bar{\phi}_a(x;\alpha)$,
simply differentiate both sides of the static field equations with
respect to the parameter:
\[
\frac{d^2\bar{\phi}_a}{dx^2}(x;\alpha)=\frac{\partial
V}{\partial\phi_a}\left |_{\vec{\bar{\phi}}}\,\,(x;\alpha) \right.
\qquad \Rightarrow \qquad
\frac{d^2}{dx^2}\cdot\frac{\partial\bar{\phi}_a}{\partial\alpha}(x;\alpha)=\sum_{b=1}^3
\frac{\partial^2V}{\partial\phi_a\partial\phi_b}\left
|_{\vec{\bar{\phi}}}\,\,(x;\alpha)\cdot\frac{\partial\bar{\phi}_a}{\partial\alpha}(x;\alpha)
\right.\qquad .
\]

On the other hand, the spectral problem is solvable for a
particular member, the $K_3^{OD}(0,0)$ kink, living in the
$\phi_1$ axis. The operator is diagonal with entries
\begin{eqnarray*}
{\cal H}_{11} &=&-\frac{d^2}{dy^2}+5-3 \tanh \sqrt{2}
y- \frac{15}{2} \, {\rm sech}^2 \sqrt{2} y \\
{\cal H}_{22} &=&-\frac{d^2}{dy^2}+1-2 \sigma_2^2+2 \sigma_2^4 +
(1-2 \sigma_2^2)\tanh \sqrt{2} y-\frac{3}{2} \, {\rm sech}^2
\sqrt{2} y  \\
{\cal H}_{33} &=&-\frac{d^2}{dy^2}+1-2 \sigma_3^2+2 \sigma_3^4 +
(1-2 \sigma_3^2)\tanh \sqrt{2} y-\frac{3}{2} \, {\rm sech}^2
\sqrt{2} y
\end{eqnarray*}
where $y=x+\frac{1}{\sqrt{2}}\log \frac{\sigma_2
\sigma_3}{\bar\sigma_2 \bar\sigma_3}$. ${\cal H}_{11}$ governs the
behavior of the fluctuations tangent to the orbit on the solution
whereas ${\cal H}_{22}$ and  ${\cal H}_{33}$ govern the orthogonal
perturbations to the kink orbit, pushing the solution towards
another with non-null values of $a$ and $b$ . The discrete
spectrum of ${\cal H}_{11}$ comprises only a zero mode whose
eigenfunction is $\frac{\partial \bar\phi}{\partial x}$, whereas
the continuous spectrum $\omega^2=q^2+1$ is non-degenerate for
$\omega^2 \in [1,4]$ and doubly degenerate for values greater than
$4$. Likewise ${\cal H}_{22}$ presents a zero mode with
eigenfunction $\frac{\partial \bar\phi}{\partial a}$ and the
continuous spectrum is non degenerate in the range
$[2\sigma_2^4,2\bar\sigma_2^4]$; doubly degeneracy of the spectrum
starts from this value. Finally, ${\cal H}_{33}$ holds a zero mode
with eigenfunction $\frac{\partial \phi}{\partial b}$ and
$[2\sigma_3^4,2\bar\sigma_3^4]$ and $(2 \max
\{\sigma_3^4,\bar\sigma_3^4\},\infty)$ are respectively the ranges
of the singly and doubly degenerate continuous spectrum. We
conclude the stability of this solution.

\subsection{Instability of four-body, six-body and seven-body solitary waves}

Note that the $K_4^{BB}(b)$ and $K_4^{CC}(b)$ kink trajectories
always meet at a common point: the conjugate point, see Figure
11(c) and 12(c). For the $K_4^{BB}$ family, this point is the
focus $F_1$, whereas for the family $K_4^{CC}$ it is one of
umbilicus points $A$ of ${\mathbb E}$. This is the signal of the
instability of these solutions: according to the Morse index
theorem, the Hessian for any member of these families has a
negative eigenvalue \cite{Ito2,J1}. The undefined flow at the foci
of the ellipse $e_1$ or the umbilical points of the ellipsoid
${\mathbb E}$ arise because the sign combinations in the Hamilton
characteristic function providing these trajectories are such
that, back in Cartesian coordinates, $W$ is a function which is
non-differentiable at either $F_1$ or $A$. Thus, these are flow
lines obtained by gluing two flow lines generated by different
Hamilton characteristic functions  at these points . The energy
depends not only on the value of the Hamilton function at the
beginning and the end of the orbit but also on the values at the
conjugate points. The absolute minimum value of the energy in the
corresponding topological sector is non-saturated and four-body
kinks are unstable.

For the one-parametric six-body $K_6^{CC}(b)$ and $K_6^{BB}(b)$
kink families, the situation is exactly the same. There exists a
conjugate point, which is crossed by all the members of each
family. For the $K_6^{CC}(b)$ it is the focus $F_2$ of the ellipse
$e_2$, and for the $K_6^{BB}(b)$ it is the focus $F_3$ of the
ellipse $e_3$. Again, application of the Morse index theorem and
considerations on the impossibility of reaching the absolute
minimum of the energy when there are conjugate points reveal the
instability of six-body lumps.

The case of seven-body solitary waves
$K_7^{BC}(\gamma_2,\gamma_3)$ is more complicated. There exist
focal lines, curves of conjugate points; on each point of the
characteristic hyperbola $h$ a whole one-parametric subfamily of
trajectories meet. The same is true for a point in the
characteristic ellipse $e_4$. The Morse index theorem tells us
that the Hessian for these solutions has two negative eigenvalues.
Furthermore, the absolute minimum of the energy is not reached by
$K_7^{BC}(\gamma_2,\gamma_3)$ kinks because of the existence of
lines of conjugate points crossed by the kink orbits: the
instability of seven-body solitary waves is proved.

The dynamical reason for the instability of these solitary waves
is the peculiar configuration of the lumps arranged in these
solutions. All these solitary waves are static solutions formed by
several basic particles, where some of these lumps travel
superposed. Because they are static solutions, the lumps are
distributed in order to achieve an equilibrium configuration where
no forces between lumps are involved: the exact overlapping
between lumps produces this balance of forces. Any small
perturbation slightly splitting the superposed lumps breaks the
unstable equilibrium releasing forces between the lumps. The
negative eigenvalues of the Hessian operator associated with this
process indicate that the evolution of these solitary waves moves
them away from the initial static configurations.

\section{Three-body low-energy scattering of non-linear waves}

The three parameters $x_0$, $a$ and $b$ of the $K_3^{OD}$ family
of solitary waves fix the center of mass and the relative
positions of the basic lumps in the composite kink configuration.
We shall now study the evolution of composite solitary waves
within the framework of Manton's adiabatic principle, see
\cite{Aai2, Manton}, by looking at changes in $a$ and $b$ in time.
The Manton adiabatic hypothesis can be summarized as follows: the
dependence on time of solitary waves develops only in the
parameters of the solution, i.e. $\bar\phi^{\rm
K}(x;t)=\bar\phi^{\rm K}(x;a(t),b(t))$. This is a good
approximation if the kink evolution is so slow that the
differential equations --formula (5) in part I-- are satisfied at
every given $t$ with good accuracy.

The most general family of trajectories arising in a
$N$-dimensional Hamiltonian system depends on $2N$ parameters: $N$
separation constants and $N$ integration constants. Finite action
fixes the separation constants at a fixed value. Therefore, the
most general family of solitary waves in $N$-component scalar
field theory depends on $N$ parameters, the integration constants
of the finite action trajectories in the analogous mechanical
system. In this case adiabatic evolution is of the general form:
\[
\bar\phi_a^{\rm K}(x;t)=\bar\phi^{\rm K}(x;a^I(t)) \qquad ,
\qquad  a=1,2, \cdots , N \qquad , \qquad I=1,2, \cdots , N
\]
Introducing this assumption in the action functional,
(\ref{eq:action}) one obtains the action for a geodesic problem:
\[
S^G=\frac{1}{2} \int dx \,dt \sum_{a=1}^N \frac{\partial
\bar\phi_a^{\rm K}}{\partial t} \frac{\partial \bar\phi_a^{\rm
K}}{\partial t} = \frac{1}{2} \int dt \sum_{I,J=1}^N g_{IJ} (a^K)
\frac{d a^I}{dt}\frac{d a^J}{dt} \qquad ,
\]
where the metric tensor $g_{IJ}(a^K)$, the Christoffel symbols
$\Gamma_{IJ}^K$, and the curvature tensor are:
\[
g_{IJ}(a^K)=\sum_{a=1}^N \int_{-\infty}^\infty dx\, \frac{\partial
\bar\phi_a^{\rm K}}{\partial a^I} \frac{\partial \bar\phi_a^{\rm
K}}{\partial a^J} \qquad , \qquad \sum_{K=1}^N
g_{MK}\Gamma_{IJ}^K=\sum_{i=1}^N \int_{-\infty}^\infty dx
\frac{\partial \bar{\phi}_i}{\partial a^M} \frac{\partial^2
\bar{\phi}_i}{\partial a^I \partial a^J}
\]
\[
R_{IJKM}=\sum_{i=1}^n \int_{-\infty}^\infty dx \left[
\frac{\partial^2 \bar\phi_i^{\rm K}}{\partial a^I \partial a^K}
\frac{\partial^2 \bar\phi_i^{\rm K}}{\partial a^M \partial a^J}
-\frac{\partial^2 \bar\phi_i^{\rm K}}{\partial a^I \partial a^M}
\frac{\partial^2 \bar\phi_i^{\rm K}}{\partial a^K \partial a^J}
\right] \quad , \quad R_{IJKM}+R_{IKMJ}+R_{IMJK}=0 \qquad ,
\]
\[
R_{IJKM}=-R_{JIKM}=-R_{IJMK}=R_{JIMK}=R_{KMIJ} \qquad .
\]

Study of the evolution of solitary waves thus turns into a
geometric problem in the adiabatic regime: the analysis of
geodesic motion in the moduli space of parameters where a
non-Euclidean metric is inherited from the dynamics of the zero
modes. The geodesics obey the associated Lagrange equations:
\begin{equation}
\frac{d^2 a^K}{dt^2}+\sum_{I,J=1}^n \Gamma_{IJ}^K
\frac{da^I}{dt}\frac{da^J}{dt}=0 \hspace{0.2cm}; \hspace{0.2cm}
K=1,\dots, N \label{eq:geode}
\end{equation}

Back in our $N=3$ scalar field theory we consider $a^{I=1}=a$ and
$a^{I=2}=b$ as local coordinates in the moduli space of parameters
for $K_3^{OD}(a,b)$ three-body solitary waves. Both the dynamics
and the geometry of the center of mass determined by the $x_0$
parameter are trivial: the component $g_{x_0 x_0}$ is constant,
the geodesics being straight lines $x_0={\rm v}\, t+d$. Therefore,
the dynamics of the center of mass obeys Galilean, rather than
Lorentzian, velocity transformations in the adiabatic regime.
Choosing the values of the coupling constants
$\sigma_3^2=2\sigma_2^2=\frac{1}{3}$ and defining the variable
$z=e^{\frac{2\sqrt{2}x}{3}}$ as in Section \S 3, we obtain the
following integral expressions for the components of the metric
tensor:
\begin{eqnarray}
g_{aa}(a,b)&=& \int_0^\infty
\frac{2(1+b^2z^2+z^3)^2+a^2(3z+b^2
z^3)}{2\sqrt{2} (1+a^2 z+b^2 z^2+z^3)^3} dz \label{eq:met1}\\
g_{ab}(a,b)&=& \int_0^\infty \frac{-a b z^3 (a^2+2 b^2 z+3
z^2)}{2\sqrt{2} (1+a^2 z+b^2 z^2+z^3)^3} dz \label{eq;met2}\\
g_{bb}(a,b)&=& \int_0^\infty \frac{z[(a^4+3b^2)z^2+(1+z^3)^2+2a^2
z (1+b^2 z^2+z^3) ]}{2\sqrt{2} (1+a^2 z+b^2 z^2+z^3)^3} dz
\label{eq:met3} \qquad .
\end{eqnarray}
Integrals of rational functions in (\ref{eq:met1}),
(\ref{eq;met2}) and (\ref{eq:met3}) would be feasible but,
depending on the arbitrary parameters $a$ and $b$, the quadratures
involved in this calculation lead to expressions that are so
complicated that there is no advantage in addressing the problem
analytically. Numerical integration, however, gives us interesting
qualitative information about the metric tensor and the scalar
curvature. In Figure 16, the components of the metric tensor as
well as the scalar curvature are displayed as a function of $a$
and $b$.

\setcounter{figure}{15}

\begin{figure}[hbt]
\centerline{
\includegraphics[height=3cm]{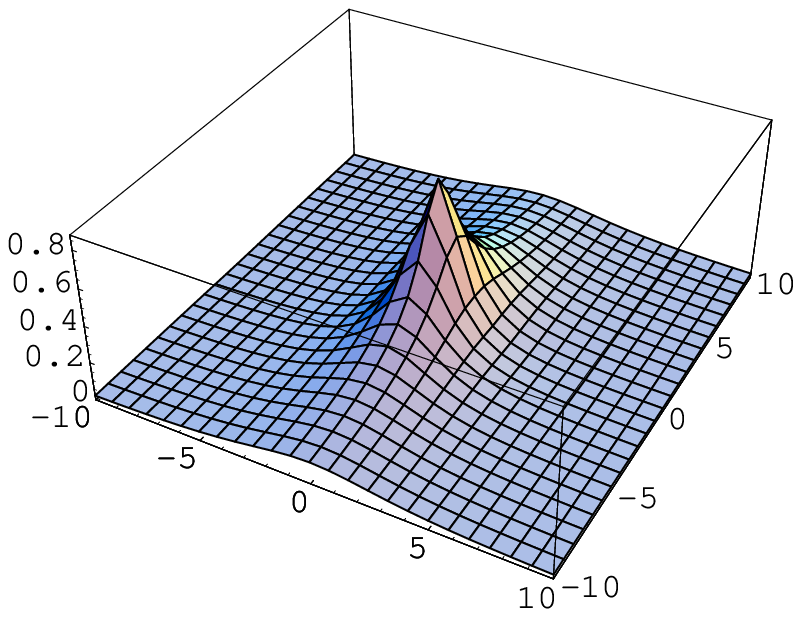}
\includegraphics[height=3cm]{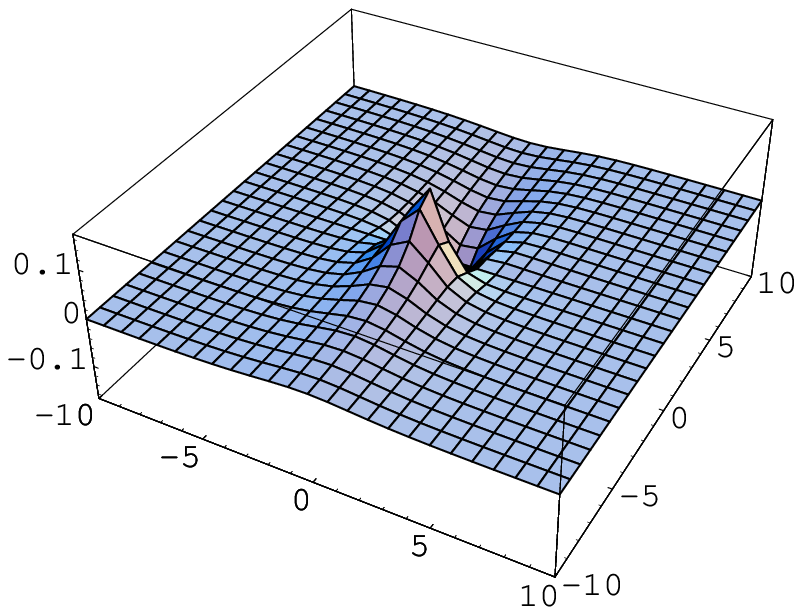}
\includegraphics[height=3cm]{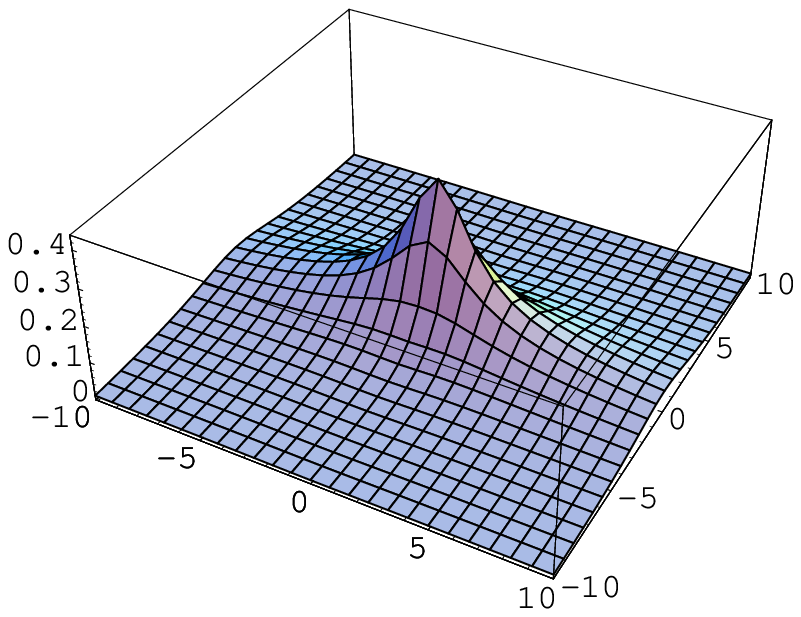}
\includegraphics[height=3cm]{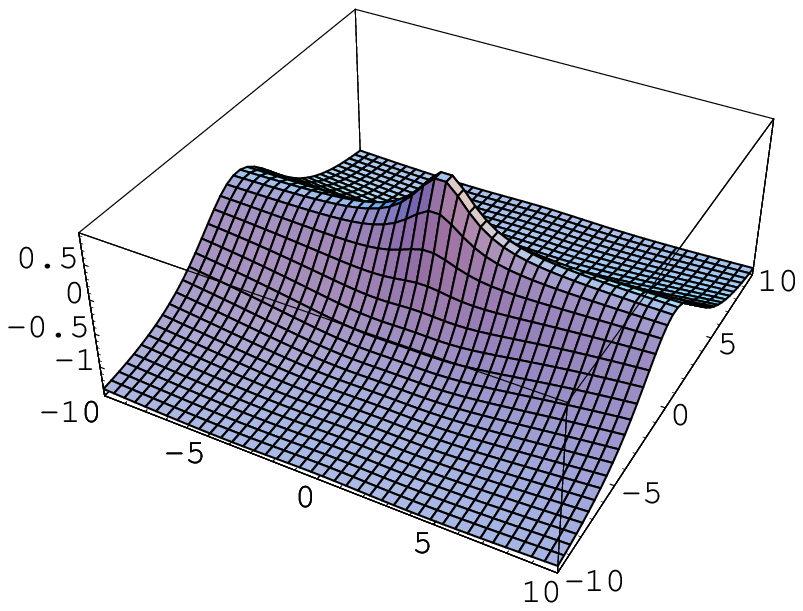}
} \caption{\small \textit{Metric Components and Scalar Curvature:
a) $g_{aa}(a,b)$, b) $g_{ab}(a,b)$, c) $g_{bb}(a,b)$, d) R(a,b).
}}
\end{figure}
The picture of the scalar curvature is especially useful for
obtaining clues about geodesic motion. In regions where the
curvature is positive the geodesics tend to approach each other
whereas where it is negative the geodesics move away. Therefore, a
certain interaction between lumps arises in the study of the
evolution of solitary waves even in the adiabatic approximation.
Recall that $K_3^{OD}(x;a,b)$ solitary waves are stable and the
zero modes are the lowest energy eigenfunctions of the Hessian
operator; low-energy fluctuations are dominated mainly by these
channels, certifying the efficacy of the adiabatic procedure. For
this reason, the adiabatic method is not applicable to unstable
configurations, such as four-body, $K_4^{BB}(b)$ or $K_4^{CC}(b)$,
six-body, $K_6^{CC}(b)$ and $K_6^{BB}(b)$, and seven-body,
$K_7^{BC}(a,b)$, solitary waves, because negative eigenvalues
arise in the spectrum of the Hessian for these kinks.

To gain more precise knowledge about adiabatic motion we
numerically solve the geodesic equation (\ref{eq:geode}) for some
different choices of the initial conditions $a(0)$, $b(0)$,
$\dot{a}(0)$ and $\dot{b}(0)$. The values of the parameters $a(t)$
and $b(t)$ at $t=0$ specify the initial static configuration
$K_3^{OD}(a,b)$ and the soft perturbations on the kinks due to the
zero modes $\frac{\partial \phi}{\partial a}$ and $\frac{\partial
\phi}{\partial b}$ are determined by the values of the
time-derivatives $\dot{a}(t)$ and $\dot{b}(t)$ at $t=0$.

We first describe the scattering process when a perturbation due
to the zero mode $\frac{\partial \phi}{\partial b}$ is exerted on
a $K_3(0,b)$ kink with $|b|$ large.  The initial solitary wave
starts from some point in the green zone C of the moduli space
(Figure 9) so that the configuration is formed by two lumps: a
$K_1^{OB}$ kink is on the right of the energy density profile and
the superposition of a $K_1^{BC}$ and a $K_1^{CD}$ kink is on the
left. These lumps approach each other, later crash into each other
in the red zone A, and finally bounce back, recovering their
original shape again in the other branch of green zone C, see
Figure 17(a).

The process illustrated in Figure 17(b) is similar to the previous
one. Here, a perturbation due to the zero mode $\frac{\partial
\phi}{\partial a}$ is applied to a $K_3(a,0)$ solitary wave. The
initial configuration is formed by a $K_1^{CD}$ lump and the
superposition of a $K_1^{BC}$ and a $K_1^{OB}$ kink. The process
of scattering is as follows: the two lumps approach each other in
the green zone B, collide and travel superposed along the red zone
A for a period of time and finally split into two lumps,
recovering the initial configuration in the other branch of green
zone B.

\begin{figure}[hbt]
\begin{tabular}{l}\includegraphics[height=4.5cm]{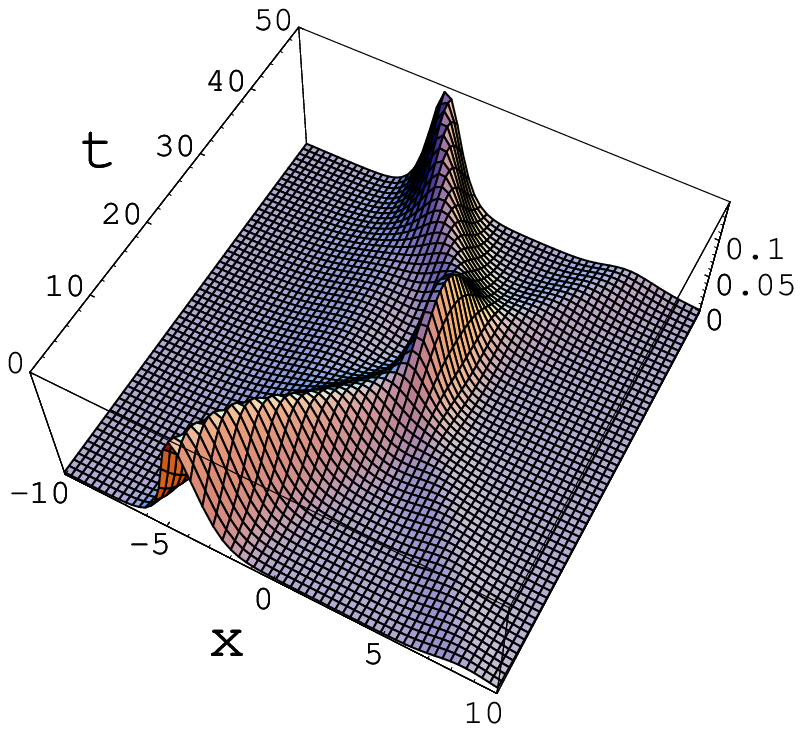}\end{tabular}
\begin{tabular}{l}
\includegraphics[height=1.5cm]{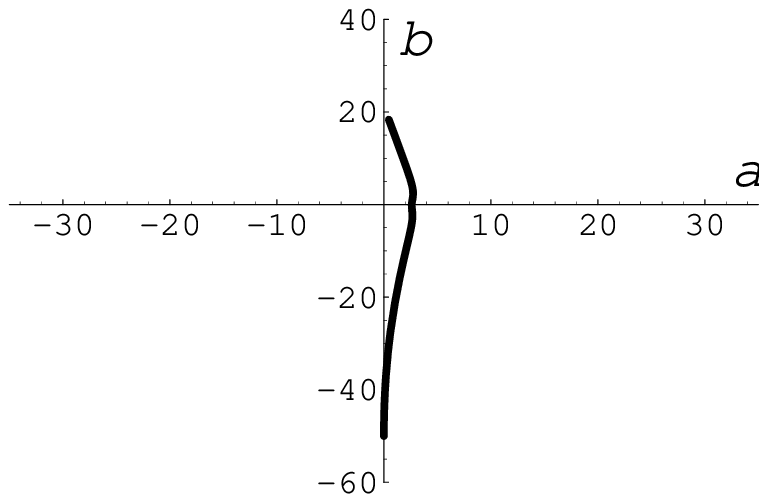} \\
\includegraphics[height=1.5cm]{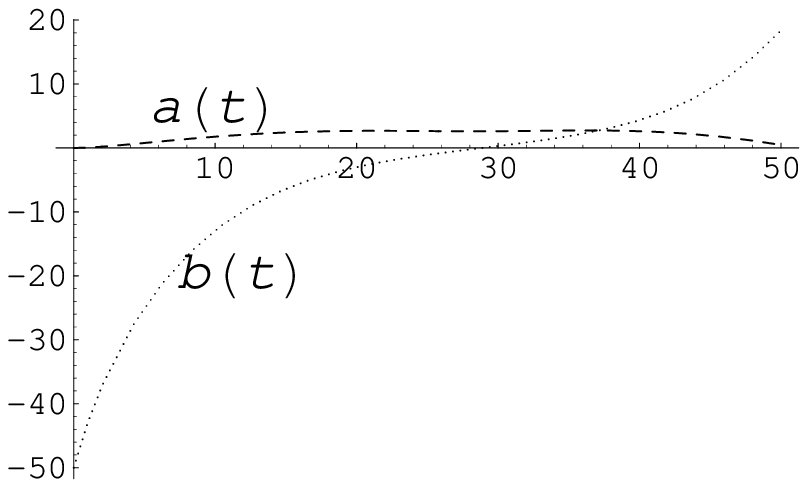}
\end{tabular}
\begin{tabular}{l}\includegraphics[height=4.5cm]{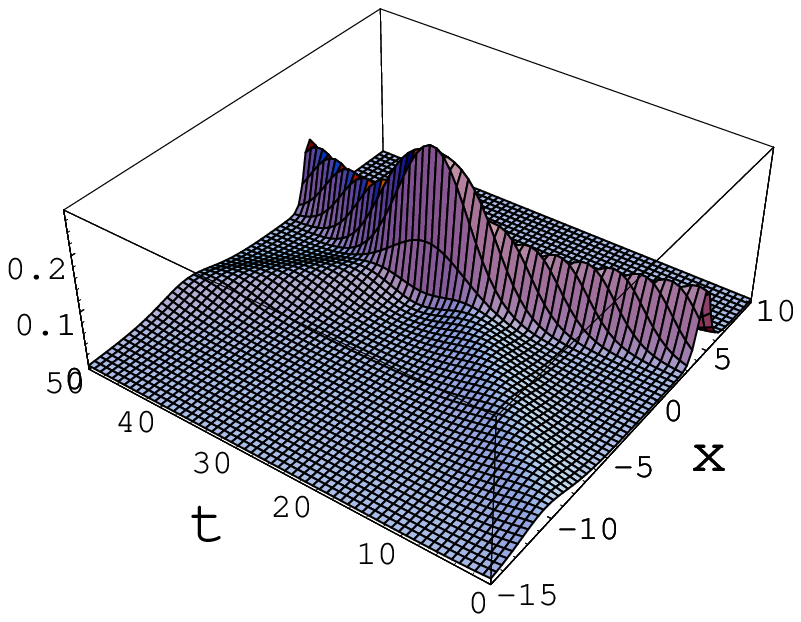}\end{tabular}
\begin{tabular}{l}
\includegraphics[height=1.5cm]{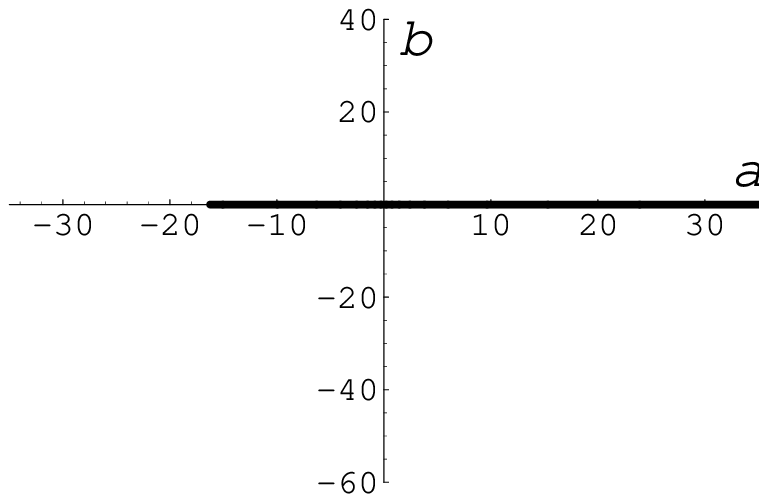} \\
\includegraphics[height=1.5cm]{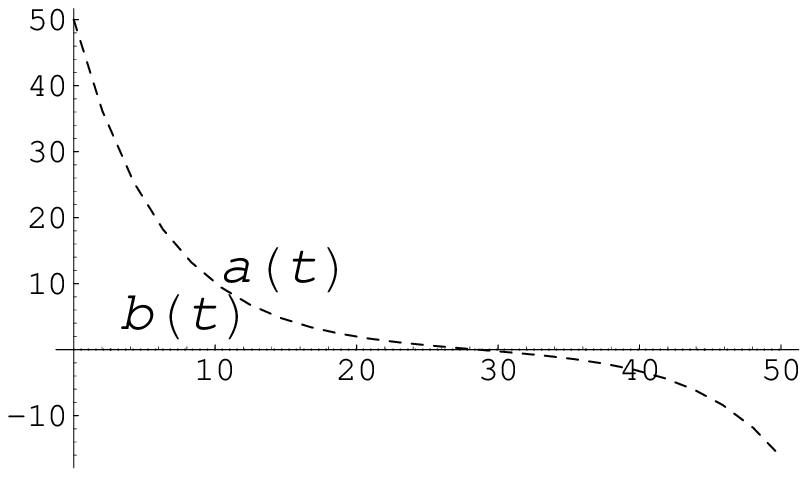} \end{tabular} \caption{\small
\textit{Evolution of solitary waves. }}
\end{figure}

Although one of the two lumps is a composite of two basic
particles, the phenomena described above are two-body scattering
processes, essentially identical to the collisions of solitary
waves studied in \cite{Mar} and \cite{modeloa}. More sophisticated
scattering processes are displayed in Figure 18, triggered by
perturbations due to the two zero modes . {\it Genuine three-body
low energy scattering arises, and as a result the main novelty in
this paper emerges}.

First, start from a $K_3^{OD}(10,-70)$ solitary wave in the blue
zone D near the border with the green zone C.  In this
configuration, the $K_1^{CD}$ and $K_1^{BC}$ kinks are close but
the $K_1^{OB}$ lump is far away from the other two, see Figure
18(a). A perturbation characterized by the values $\dot{b}(0)=-3$,
$\dot{a}(0)=6$ is exerted on this arrangement of lumps. The
evolution of the solitary wave -unveiled from the corresponding
geodesic found numerically- is as follows: the two closer
particles collide and merge whereas the other particle goes
towards them, all entering together in green zone C. The united
lumps travel together and eventually split into the initial basic
lumps -back again in a blue zone D of moduli space-, the
$K_1^{CD}$ moving towards the left and the $K_1^{BC}$ towards the
right on the line. Later, this lump bumps into the $K_1^{OB}$
kink, which was traveling towards the left. The $K_1^{BC}$ and the
$K_1^{OB}$ basic particles merge and move as a single one, -now in
a green zone B-, and, after a certain time, they split, the whole
solitary wave again entering in other blue zone D in the moduli
space. Then, the $K_1^{OB}$ kink advances towards the right and
the $K_1^{BC}$ advances towards the left on the line. Now this
particle is attracted by the $K_1^{CD}$ lump and they merge,
coming again to a green zone C. In conclusion, the process can be
interpreted as a interchange of the particle $K_1^{BC}$ between
the other particles $K_1^{CD}$ and $K_1^{OB}$.

A last example of three-body collisions is displayed in Figure
18(b). A perturbation determined by $\dot{b}(0)=-3$,
$\dot{a}(0)=6$ is applied on a initial $K_3^{OD}(50,-50)$ solitary
wave living in a blue zone D of the moduli space. The
$K_3^{OD}(50,-50)$ configuration is formed by three separated
basic particles. Note the apparent attractive forces between the
lumps in Figure 18(b) showing the evolution of this solitary wave
when the perturbation chosen above acts on it. The basic particles
tend to approach each other at the beginning of the geodesic
motion. More precisely, first the $K_1^{OB}$ and $K_1^{BC}$ lumps
amalgamate when the geodesic enters a green zone B of the moduli
space, then they travel together and, later on, they all merge
together with the $K_1^{CD}$ kink. During a period of time, when
the geodesic passes through the red zone A, the three particles
move together and, finally, a process of fission takes place in
two stages. The $K_1^{CD}$ kink splits from the global lump when
the geodesic leaves the red zone to enter into the other green
zone B, and later on so do the $K_1^{OB}$ and $K_1^{BC}$ kinks,
the geodesic ending back in other blue zone D. This process may be
repeated if the initial conditions setting the magnitude of the
perturbation and the initial solitary wave are chosen adequately.

\begin{figure}[hbt]
\begin{tabular}{l}\includegraphics[height=4.5cm]{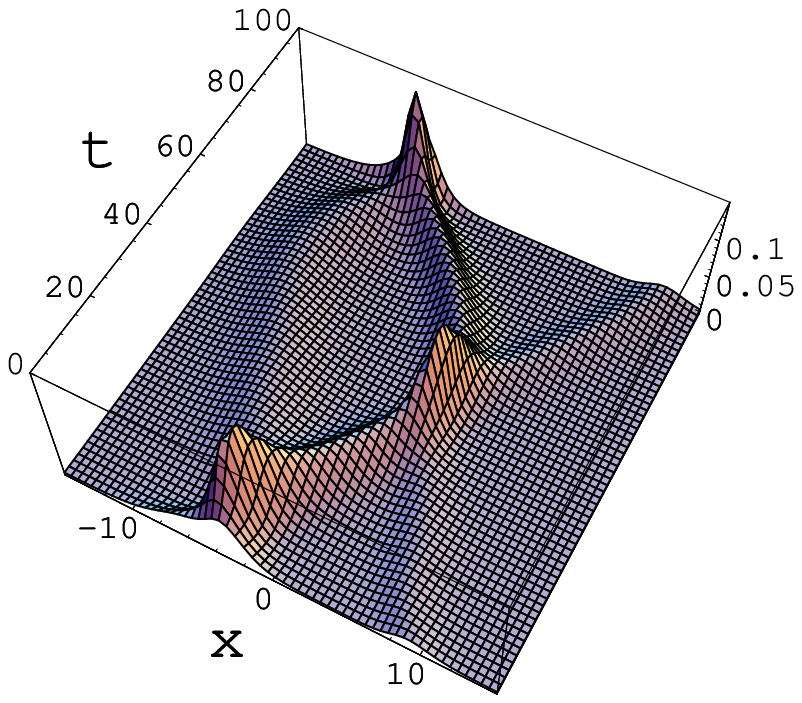}\end{tabular}
\begin{tabular}{l}
\includegraphics[height=1.5cm]{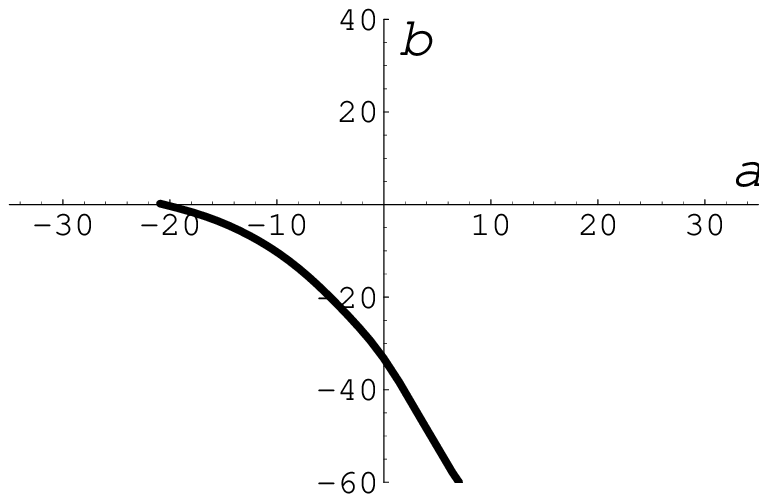} \\
\includegraphics[height=1.5cm]{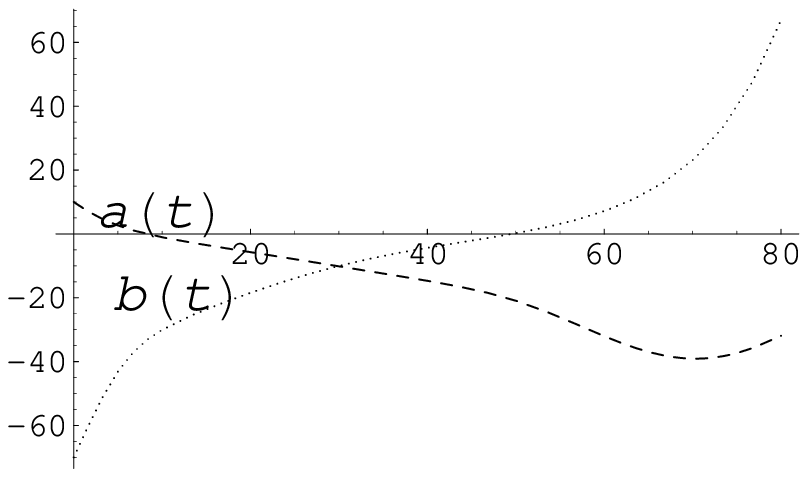}
\end{tabular}
\begin{tabular}{l}\includegraphics[height=4.5cm]{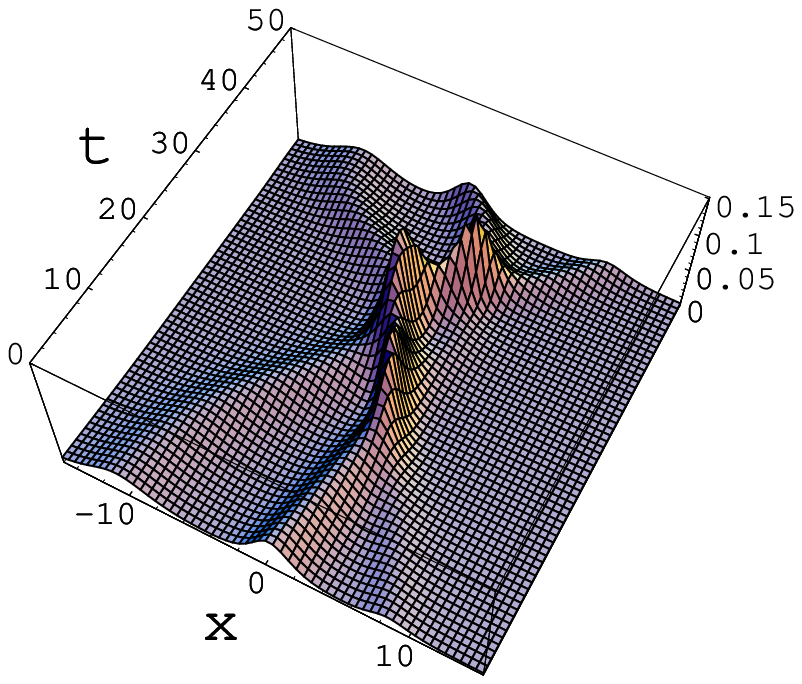}\end{tabular}
\begin{tabular}{l}
\includegraphics[height=1.5cm]{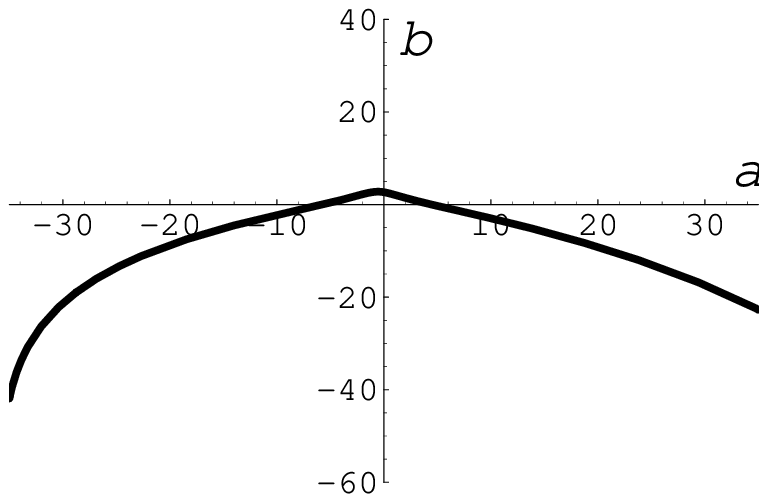} \\
\includegraphics[height=1.5cm]{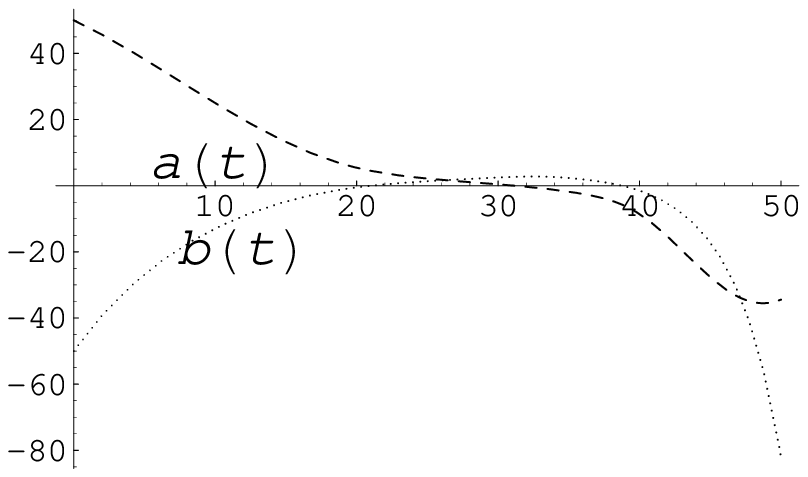} \end{tabular} \caption{\small
\textit{Evolution of solitary waves. }}
\end{figure}

\end{document}